\documentclass[prd,nofootinbib,showkeys]{revtex4}
\pdfoutput=1
\usepackage{graphicx}
\usepackage[a4paper, hdivide={3cm,,3cm}, vdivide={2.5cm,,3cm}]{geometry}
\usepackage{amstext,amssymb}
\usepackage[intlimits]{amsmath}
\usepackage[english]{babel}
\usepackage[ansinew]{inputenc}
\usepackage{units}
\usepackage{dsfont} % just for a nice identity matrix \mathds{1}
\usepackage{hyperref}

\hypersetup{
    pdftitle={"Hidden" O(2) and SO(2) Symmetry in Lepton Mixing},    % title
    pdfauthor={Julian Heeck, Werner Rodejohann}     % author
}

\begin{document}
\let \Lold \L
\def \L {\mathcal{L}} %Lagrangian density
\let \epsilonold \epsilon
\def \epsilon {\varepsilon} %different epsilon symbol
\let \arrowvec \vec
\def \vec#1{{\boldsymbol{#1}}}
\def\ra{\rightarrow}
\newcommand{\matrixx}[1]{\begin{pmatrix} #1 \end{pmatrix}} %Matrix with brackets
\def\2vec#1#2{ \matrixx{#1 \\ #2} } %nice looking two-vector
\newcommand{\hc}{\mathrm{h.c.}}
\newcommand{\diag}{\mathrm{diag}}
\def \Z#1{{\mathbb{Z}_{#1}}} % discrete Z_N symmetry \Z{N}
\def \eins {\mathds{1}} % cute identity matrix
\newcommand{\datm}{\Delta m_\mathrm{atm}^2} % atm. mass squared
\newcommand{\dsol}{\Delta m_\mathrm{sol}^2} % sol. mass squared

\title{``Hidden'' $\boldsymbol{O(2)}$ and $\boldsymbol{SO(2)}$ Symmetry in Lepton Mixing}
\author{Julian \surname{Heeck}}
\email{julian.heeck@mpi-hd.mpg.de}
\affiliation{Max--Planck--Institut f\"ur Kernphysik,\\Saupfercheckweg 1, 69117 Heidelberg, Germany}

\author{Werner \surname{Rodejohann}}
\email{werner.rodejohann@mpi-hd.mpg.de}
\affiliation{Max--Planck--Institut f\"ur Kernphysik,\\Saupfercheckweg 1, 69117 Heidelberg, Germany}
\begin{abstract}
To generate the minimal neutrino Majorana mass matrix that has a free solar mixing angle and $\dsol = 0$ it suffices to implement an $O(2)$ symmetry, or one of its subgroups $SO(2)$, $\Z{N\geq 3}$, or $D_{N\geq 3}$. This $O(2)$ generalizes the hidden $\mathbb{Z}_2^s$ of lepton mixing and leads in addition automatically to $\mu$--$\tau$ symmetry. Flavor-democratic perturbations, as expected e.g.~from the Planck scale, then result in tri-bimaximal mixing. We present a minimal model with three Higgs doublets implementing a type-I seesaw mechanism with a spontaneous breakdown of the symmetry, leading to large $\theta_{13}$ and small $\dsol$ due to the particular decomposition of the perturbations under $\mu$--$\tau$ symmetry.
\end{abstract}
\maketitle

%%%%%%%%%%%%%%%%%%%%%%%%%%%%%%%%%%%%%%%%%%%%%%%%%%%%%%%%%%%%%%%%%%%%%%%%%%%%%%%%%%%%%%%%%%%%%%%%%%%%%%%%%%%%%%%%%%%%%%%%

\section{Introduction}
Several decades worth of neutrino experiments have shown that at least two neutrinos are massive---sub-eV---and are subject to substantial mixing. The Pontecorvo-Maki-Nakagawa-Sakata (PMNS) leptonic mixing matrix that connects flavor- and mass-eigenstates can be parameterized as a product of three unitary matrices $U_{ij}$ and a diagonal phase matrix in the form $U_\mathrm{PMNS} = U_{23} U_{13} U_{12}\, \diag (1,  e^{i\alpha}, e^{i\beta})$, or, explicitly in components (with $c_{ij}\equiv \cos \theta_{ij}$ and $s_{ij}\equiv \sin \theta_{ij}$):
\begin{align}
\begin{split}
	U_\mathrm{PMNS} &= 
\matrixx{1 & 0 & 0 \\ 0 & c_{23} & s_{23} \\ 0 & -s_{23} & c_{23}}
\matrixx{c_{13} & 0 & e^{-i\delta_\mathrm{CP}} s_{13} \\ 0 & 1 & 0 \\ -e^{i\delta_\mathrm{CP}} s_{13} & 0 & c_{13}}
\matrixx{c_{12} & s_{12} & 0 \\ -s_{12} & c_{12} & 0 \\ 0 & 0 & 1} 
\diag (1,  e^{i\alpha}, e^{i\beta})\\
&=
\matrixx{c_{12} c_{13} & s_{12} c_{13} & s_{13} e^{-i\delta_\mathrm{CP}}\\
	-c_{23} s_{12}- s_{23} s_{13} c_{12} e^{i\delta_\mathrm{CP}} & c_{23} c_{12}- s_{23} s_{13} s_{12} e^{i\delta_\mathrm{CP}} & s_{23} c_{13}\\
	s_{23}s_{12}- c_{23} s_{13} c_{12} e^{i\delta_\mathrm{CP}} & -s_{23} c_{12}- c_{23} s_{13} s_{12} e^{i\delta_\mathrm{CP}} & c_{23} c_{13}}  
\diag (1,  e^{i\alpha}, e^{i\beta})\, ,
\end{split}
\label{eq:mixing_matrix_pmns}
\end{align}
in a basis where the charged lepton mass matrix is diagonal. The
CP-violating Majorana phases $\alpha$ and $\beta$ are not observable
in oscillation experiments and the Dirac phase $\delta_\mathrm{CP}$
only for nonvanishing reactor angle $\theta_{13}\neq 0$. 
The current best-fit values and confidence levels for the mass-squared
differences and mixing angles are taken from Ref.~\cite{theta13} and listed in
Tab.~\ref{tab:mixingangles}. We also quote the $95\%$~C.L.~range for
$\theta_{13}$ from an analysis including the Double Chooz (DC) result
for normal mass ordering (inverted mass ordering)~\cite{Machado}:
\begin{align}
 0.023 \ (0.027) < \sin^2 2\theta_{13} < 0.16 \ (0.17) \,.
\label{eq:t13boundDC}
\end{align}

\begin{table}[tb]
   \begin{tabular}{|c|c|c|c|}
       \hline
       parameter & best-fit$^{+1\sigma}_{-1\sigma}$ & 2$\sigma$ & 3$\sigma$
       \\
       \hline
       $\Delta m^2_{21}\: [10^{-5}\unit{eV^2}]$
& $7.59^{+0.20}_{-0.18}$ & 7.24--7.99 & 7.09--8.19 \\[1mm] %
$\Delta m^2_{31}\: [10^{-3}\unit{eV^2}]$
&
   \begin{tabular}{c}
     $2.50^{+0.09}_{-0.16}$\\
     $-(2.40^{+0.08}_{-0.09})$
   \end{tabular}
&
   \begin{tabular}{c}
     $2.25$--$2.68$\\
     $-(2.23$--$2.58)$
   \end{tabular}
&
   \begin{tabular}{c}
     $2.14$--$2.76$\\
     $-(2.13$--$2.67)$
   \end{tabular} \\[4mm] %% 
   $\sin^2\theta_{12}$
& $0.312^{+0.017}_{-0.015}$ & 0.28--0.35 & 0.27--0.36\\[1mm]
       $\sin^2\theta_{23}$
&
   \begin{tabular}{c}
     $0.52^{+0.06}_{-0.07}$\\
     $0.52^{+0.06}_{-0.06}$
   \end{tabular}
&
   \begin{tabular}{c}
         0.41--0.61\\
         0.42--0.61
       \end{tabular}
       & 0.39--0.64 \\[4mm] 
       $\sin^2\theta_{13}$
&
   \begin{tabular}{c}
         $0.013^{+0.007}_{-0.005}$\\
     $0.016^{+0.008}_{-0.006}$
   \end{tabular}
&
   \begin{tabular}{c}
    0.004--0.028\\
        0.005--0.031
   \end{tabular}
&
   \begin{tabular}{c}
         0.001--0.035\\ 
         0.001--0.039
   \end{tabular}\\
       \hline
\end{tabular}
\caption{ \label{tab:mixingangles} Current neutrino oscillation
parameters from global fits, taken from Ref.~\cite{theta13}. The upper
(lower) row corresponds to normal (inverted) neutrino mass hierarchy,
with $\Delta m^2_{31}>0$ ($\Delta m^2_{31}<0$). 
Recent Double Chooz results~\cite{doublechooz} are not included.}
\end{table}

The closeness of $\theta_{13}$ to zero and the atmospheric mixing angle $\theta_{23}$ to $\pi/4$ has spawned a lot of interest in models that predict these values by invoking some symmetry. Since $\theta_{12}$ also approximately fulfills $\sin^2\theta_{12} = 1/3$, the most-studied approximation to $U_\mathrm{PMNS}$ is given by so-called tri-bimaximal mixing (TBM)~\cite{tbm}
\begin{equation}
	U_\mathrm{TBM} = \matrixx{\sqrt{2/3} & 1/\sqrt{3} & 0\\
-1/\sqrt{6} & 1/\sqrt{3} & -1/\sqrt{2} \\ -1/\sqrt{6} & 1/\sqrt{3} &
1/\sqrt{2}} 
\Rightarrow \begin{array}{c}
\sin^2 \theta_{13} = 0 \times \cos^2 \theta_{13} = 0 \\
\sin^2 \theta_{12} = \frac 12 \times \cos^2 \theta_{12} = \frac 13 \\
\sin^2 \theta_{23} = 1 \times \cos^2 \theta_{23} = \frac 12 
\end{array} \, , 
\label{eq:TBM}
\end{equation}
where we choose a different (physically equivalent) sign convention
for $\theta_{23}$ compared to Eq.~\eqref{eq:mixing_matrix_pmns}. While
recent T2K~\cite{t2k} and Double~Chooz~\cite{doublechooz} results
indicate $\theta_{13}\neq 0$~\cite{theta13,globalfit2},
Eq.~\eqref{eq:TBM} can still be viewed as a good leading order
approximation. The entries of $U_\mathrm{TBM}$ are reminiscent of Clebsch-Gordan coefficients, so it is not surprising that the TBM structure can be implemented by invoking symmetry groups.
This is by no means a simple endeavor but by now there exist hundreds of TBM models based on discrete nonabelian symmetries such as $A_4$, $S_4$ or $\Delta (54)$ (see Refs.~\cite{discretesymmetries} for recent reviews of this subject). To obtain the solar mixing angle  $\sin^2\theta_{12} = 1/3$ one has to work a lot harder than for $\theta_{13}=0=s_{23}^2-1/2$, as the latter simply follows from the exchange symmetry $\mu\leftrightarrow \tau$ (which is just a $\Z2$ symmetry, denoted by $\mathbb{Z}_2^{\mu\tau}$). The $\mathbb{Z}_2^{\mu\tau}$ invariant symmetric Majorana mass matrix for the neutrinos in flavor space has the structure
\begin{align}
 \mathcal{M}_\nu^{\mu\tau} = \matrixx{a & b & b\\ \cdot & c & d \\ \cdot & \cdot & c}.
\label{eq:mutausymmetry}
\end{align}
The solar mixing angle is not fixed by the $\mathbb{Z}_2^{\mu\tau}$,
but given in terms of the entries of $\mathcal{M}_\nu^{\mu\tau}$ as:
\begin{align}
 \sin^2 2 \theta_{12} = \frac{ 8 b^2 }{ (a-c-d)^2 + 8 b^2} \,,
\end{align}
where we assumed $\mathcal{M}_\nu$ to be real for simplicity, as we
will do in the rest of this paper. The phenomenologically favored TBM
value $\sin^2 2\theta_{12} = 8/9$ can be obtained for $a+b = c+ d$, as
long as $b\neq 0$. This last condition is crucial, because for
$b=0=c+d-a$, the solar angle is actually free, as we will show in this
paper. The mass matrix then has an $O(2)$ symmetry, the exact
representation of which will be derived by us using two different
motivations. One of them has already been discussed in
Ref.~\cite{scooped}, where a model based on this $O(2)$ symmetry
(actually the $SO(2)$ subgroup) is constructed, introducing two
additional Higgs doublets and one triplet, i.e.~implementing a type-II
seesaw mechanism. While this model successfully transfers the $SO(2)$
structure to the neutrino mass matrix, the necessary symmetry breaking
is hardly discussed. The additional motivation for the $O(2)$ symmetry that we add here is
connected with a ``hidden $\Z2$'' symmetry of lepton mixing \cite{hiddenZ2_He,hiddenZ2_Dicus}, and
the $O(2)$ symmetry is interpreted here as a generalization of this
$\Z2$, leading in addition automatically to $\mu$--$\tau$ symmetry.
After providing a novel motivation for the $O(2)$ symmetry, we discuss the breaking of said symmetry by the most general perturbations, but with a focus on flavor-democratic effects (for example from the Planck scale) as those immediately lead to TBM. To present an explicit realization of the symmetry, we use a type-I seesaw mechanism extended by two scalar doublets.

The layout of the paper is the following: In Sec.~\ref{sec:O2symmetry} we motivate the $O(2)$ symmetry from different perspectives and identify the subgroups that lead to the desired Majorana mass matrix. Since the symmetry cannot be exact, we discuss the most general perturbations to this mass matrix in Sec.~\ref{sec:symmetrybreaking}, with a focus on the flavor-democratic structure imposed by Planck-scale perturbations. To illustrate a realization of this symmetry we build a minimal model using type-I seesaw in Sec.~\ref{sec:minimalmodel} and show that the spontaneous breakdown can generate $\theta_{13}$ in the T2K range. We conclude our work in Section~\ref{sec:conclusions}. Appendix~\ref{app:o2reps} gives a brief overview of $O(2)$ and its representations, while Appendix~\ref{app:hiddenZ2} gives an introduction to the hidden $\Z2$ associated with $\mu$--$\tau$ symmetry, as it serves as a motivation for the $O(2)$ used in our work.

%%%%%%%%%%%%%%%%%%%%%%%%%%%%%%%%%%%%%%%%%%%%%%%%%%%%%%%%%%%%%%%%%%%%%%%%%%%%%%%%%%%%%%%%%%%%%%%%%%%%%%%%%%%%%%%%%%%%%%%%

\section{\texorpdfstring{$\boldsymbol{O(2)}$}{O(2)} Symmetry}
\label{sec:O2symmetry}
In this section we will motivate and discuss an $O(2)$ symmetry
connected to the solar mixing angle $\theta_{12}$ and $\dsol = 0$. We
present two derivations to make the discussion more lucid and also
comment on relevant subgroups. We refer to App.~\ref{app:o2reps} for a
short introduction to the group $O(2)$ and its representations. 

\subsection{Connection to Hidden \texorpdfstring{$\mathbb{Z}_2^s$}{Z2}}
\label{sec:hiddenZ2}
As shown in Ref.~\cite{hiddenZ2_He,hiddenZ2_Dicus} and App.~\ref{app:hiddenZ2}, every $\mu$--$\tau$ symmetric Majorana mass matrix~\eqref{eq:mutausymmetry} is automatically invariant under a second (so-called hidden) $\Z2= \mathbb{Z}_2^s$, generated by
\begin{align}
 G (\theta_s) \equiv -\matrixx{ \cos 2\theta_s & -\sin 2\theta_s /\sqrt{2}  & -\sin 2\theta_s /\sqrt{2} \\ \cdot & -\cos^2 \theta_s & \sin^2 \theta_s \\ \cdot & \cdot & -\cos^2 \theta_s}.
\label{eq:hiddenZ2}
\end{align}
$G(\theta_s)$ acts on the flavor eigenstates $\vec{\nu}_f \equiv (\nu_e, \nu_\mu, \nu_\tau)^T$ and satisfies $G^T = G = G^{-1}$, $\det G = -1$, so it is the generator of a $\Z2$ symmetry.
The main point of our discussion is that $G(\theta)$ can also be
viewed as a reflection, i.e.~$G(\theta) \in O(2)\backslash SO(2)$. The
$O(2)$ structure becomes apparent when we go to the
basis\footnote{This basis coincidentally diagonalizes the mass matrix,
i.e.~constitutes a mass basis. We will come back to this point later on.}
\begin{align}
\vec \nu_O = U_{23}^T(-\pi/4) \,\vec \nu_f \equiv \matrixx{1 & 0 & 0
\\ 0 &\cos (-\pi/4) & \sin (-\pi/4) \\ 0 & -\sin (-\pi/4) & \cos
(-\pi/4)}^T   \matrixx{ \nu_e \\ \nu_\mu \\ \nu_\tau }= \matrixx{\nu_e
\\ (\nu_\mu+ \nu_\tau)/\sqrt{2}\\(\nu_\mu - \nu_\tau)/\sqrt{2}} , 
\end{align}
in which $G$ takes the form
\begin{align}
 G_O \equiv U_{23}^T (-\pi/4) \, G \, U_{23} (-\pi/4) = \matrixx{- \cos 2\theta & \sin 2\theta & 0\\ \sin 2\theta & \cos 2\theta & 0 \\ 0 & 0 & 1} .
\end{align}
We identify $G_O$ as the most general reflection~(see
Eq.~(\ref{eq:generalreflection})) acting on $\vec{\nu}_O \sim \vec 2 \oplus
\vec 1$, $(\nu_\mu - \nu_\tau)/\sqrt{2}$ being the singlet. For the
doublet we can choose the basis $(\nu_e, \, (\nu_\mu+
\nu_\tau)/\sqrt{2})^T \sim \vec 2$.  
The set $\{G(\theta),\,\theta\in\mathbb{R}\}$ from Eq.~\eqref{eq:hiddenZ2} forms the subset of $O(2)$ with determinant $-1$, so to construct the whole $O(2)$ we need to find the associated rotations. As a rotation can be written using two reflections, we immediately arrive at the representation in flavor space
\begin{align}
 R (\theta_s)  = G(\theta_s/2) G(0)= \matrixx{c_s & s_s/\sqrt{2} & s_s/\sqrt{2} \\ -s_s/\sqrt{2} & \cos^2 (\theta_s/2) & - \sin^2 (\theta_s/2) \\ -s_s/\sqrt{2} & - \sin^2 (\theta_s/2) & \cos^2 (\theta_s/2)} .
\label{eq:Rinflavour}
\end{align}
The rotations~\eqref{eq:Rinflavour} and reflections~\eqref{eq:hiddenZ2} span the group $O(2)$ that generalizes the hidden $\mathbb{Z}_2^s$, which is why we named it ``hidden'' $O(2)$.
As special cases we note that the reflection $G(0) = \diag (-1,1,1)$ can be interpreted as the generator of a $\mathbb{Z}_2$ connected to the electron number, i.e.~$\nu_e$ is odd under this $\Z2$ while $\nu_{\mu}$ and $\nu_\tau$ are even, as well as 
\begin{align}
 R(\pi) = -\matrixx{1 & 0 & 0 \\ 0 & 0 & 1 \\ 0 & 1 & 0} ,
\label{eq:mutausymmetrygenerator}
\end{align}
which generates the $\mu$--$\tau$ exchange symmetry $\mathbb{Z}_2^{\mu\tau}$. 

It is interesting to study neutrino mass matrices that are invariant under the action of the
$\mathbb{Z}_2^s$, i.e.~$[\mathcal{M}_\nu, G(\theta_{12})]=0$, as they
lead to testable relations among the mixing angles and the CP-phase
$\delta_\mathrm{CP}$~\cite{hiddenZ2_Dicus}. We will go one step further and
impose $[\mathcal{M}_\nu, G(\theta_s)]=0$ $\forall \theta_s$, not just
for one special value $\theta_{12}$. From $R (\theta_s)  =
G(\theta_s/2) G(0)$ it is with Eq.~(\ref{eq:a3}) obvious that this condition already implies
invariance under rotations $[\mathcal{M}_\nu, R(\theta_s)]=0$ $\forall
\theta_s$, so $[\mathcal{M}_\nu, G(\theta_s)]=0$ $\forall \theta_s$
gives an $O(2)$ invariant mass matrix:
\begin{align}
\mathcal{M}_\nu^\mathrm{deg} = \matrixx{m_1 & 0 & 0 \\ \cdot & (m_1 + m_3)/2 & (m_1 -m_3)/2 \\ \cdot & \cdot & (m_1 + m_3)/2} .
\label{eq:so2invariant_massmatrix}
\end{align}
One can easily convince oneself that invariance under the
reflection (\ref{eq:hiddenZ2}) and rotation (\ref{eq:Rinflavour})
holds, which together span the group $O(2)$.
$SO(2)$ invariance ($[\mathcal{M}_\nu, R(\theta_s)]=0$ $\forall \theta_s$) by itself results in the same Majorana matrix, so we can not distinguish between $O(2)$ and $SO(2)$ in the neutrino sector.
Due to the automatic $\mu$--$\tau$ symmetry of $\mathcal{M}_\nu^\mathrm{deg}$ we
find $\theta_{13}=0$ and $\theta_{23}=-\pi/4$, whereas the solar
mixing angle $\theta_{12}$ is undetermined because the two mass
eigenstates $\nu_e$ and $(\nu_\mu+ \nu_\tau)/\sqrt{2}$ have the same
mass $m_1$. Consequently we have $\dsol = 0$. As a special case we
note that in the limit $m_3\ra -m_1$ the matrix
$\mathcal{M}_\nu^\mathrm{deg}$ conserves not only the flavor symmetry
$L_\mu - L_\tau$, but due to the mass degeneracy even
$SU(2)_{L_\mu-L_\tau}$~\cite{SU(2)LmuLtau}. In the following we will
discuss the mass matrix $\mathcal{M}_\nu^\mathrm{deg}$, as motivated
either by $O(2)$ or $SO(2)$ invariance. The difference becomes
important only in model building when considering the charged-lepton sector (see Sec.~\ref{sec:minimalmodel}).

It is instructive to determine the smallest finite subgroups of $O(2)$, i.e.~$\Z{N}$ and $D_N$, that lead to $\mathcal{M}_\nu^\mathrm{deg}$. The abelian groups $\Z2$ and $D_2$ are generated by $R(\pi)$ and $\{R(\pi), G(0)\}$, respectively, and do not lead to the symmetry condition $(\mathcal{M}_\nu)_{11} = (\mathcal{M}_\nu)_{22} + (\mathcal{M}_\nu)_{23}$. $\Z3$ is generated by powers of
\begin{align}
 R(2\pi/3) = \frac{1}{4}\matrixx{-2 & \sqrt{6} & \sqrt{6} \\ -\sqrt{6} & 1 & -3\\ -\sqrt{6} & -3 & 1} ,
\label{eq:Z3}
\end{align}
and $[\mathcal{M}_\nu, R(2\pi/3)]=0$ already fixes $\mathcal{M}_\nu = \mathcal{M}_\nu^\mathrm{deg}$. Consequently the smallest nonabelian $O(2)$ subgroup $D_3 \cong S_3 \cong \Z{3} \rtimes \Z2$ also leads to $\mathcal{M}_\nu^\mathrm{deg}$, because it has $\Z3$ as a subgroup. The general $D_N = \Delta (2 N)$ is generated by $R(2\pi/N)$ and e.g.~$G(0)$ (any reflection really), as they satisfy the multiplication rules~\cite{discretesymmetries}
\begin{align}
 \left[ R(2\pi/N) \right]^N = \left[ G(0) \right]^2 = \left[ R(2\pi/N) G(0) \right]^2 = \eins \,.
\end{align}
One can show that $D_{N > 3}$ and $\Z{N>3}$ also fix the form~\eqref{eq:so2invariant_massmatrix}, so we conclude that the Majorana mass matrix $\mathcal{M}_\nu^\mathrm{deg} $ can be obtained by imposing $O(2)$ or $D_{N \geq 3}$ as generated by $R$ in Eq.~\eqref{eq:Rinflavour} and $G$ in Eq.~\eqref{eq:hiddenZ2}, or $SO(2)$ or $\Z{N \geq 3}$ as generated by $R$.

As far as nonsymmetric matrices go, e.g.~Dirac mass matrices, the invariance condition under $SO(2)$ or  $\Z{N}$ ($N \geq 3$) fixes the form
\begin{align}
 R^T m_D R = m_D = \matrixx{a + b & c & c \\ -c & a & b \\ -c & b & a} ,
\label{eq:invariantmD}
\end{align}
whereas invariance under $O(2)$ or $D_N$ ($N  \geq 3$) sets $c=0$ as it flips sign under reflections. For Dirac matrices we also have the option that the left-handed fields $L_\alpha$ and the right-handed fields $\ell_\alpha$ transform differently under the $SO(2)$, i.e.~$L_\alpha \ra R_{\alpha\beta} L_\beta$ and $\ell_\alpha \ra \ell_\alpha$. This fixes the form
\begin{align}
 R^T m_D  = m_D = \matrixx{0 & 0 & 0 \\ a & b & c \\ -a & -b & -c} ,
\label{eq:2plus1times1plus1plus1}
\end{align}
resulting in only one massive particle. This is not surprising as we
couple the $O(2)$ representation $\vec{2}\oplus \vec{1}$ of $L_\alpha$ to the $\vec{1}\oplus \vec{1}\oplus
\vec{1}$ representation of $\ell_\alpha$, which allows only one invariant term unless we introduce
Higgs doublets that transform under $SO(2)$ (see Sec.~\ref{sec:minimalmodel}). The coupling of $\vec{2}\oplus \vec{1}$ to $\vec{2}\oplus \vec{1}$ from Eq.~\eqref{eq:invariantmD} on the other hand allows for two invariants, a mass for the singlet and for the doublet. The form of Eq.~\eqref{eq:2plus1times1plus1plus1} does not change when extending the symmetry to $O(2)$, as long as one of the right-handed particles transforms as $\vec{1}$ (instead of $\vec{1}'$ which is odd under reflections (see App.~\ref{app:o2reps})). Once again the discrete symmetries $\Z{N}$ and $D_N$ ($N  \geq 3$) suffice to obtain Eq.~\eqref{eq:2plus1times1plus1plus1}. 

In the concrete model of Sec.~\ref{sec:minimalmodel} we will find that three nonzero charged lepton masses are much easier to obtain in an $SO(2)$ than in an $O(2)$ symmetric model. This makes of course no difference for Majorana neutrinos, at least in the exact symmetry limit.

\subsection{Vanishing \texorpdfstring{$\boldsymbol{\dsol}$}{m2**2-m1**2}}
In this section we give an entirely different motivation for our $O(2)$ symmetry.
The TBM mixing matrix~\eqref{eq:TBM} can be obtained from the neutrino mass matrix (in a basis where the charged lepton mass matrix is diagonal)
\begin{align}
\begin{split}
 \mathcal{M}_\nu^\mathrm{TBM} &= U_\mathrm{TBM} \,\diag (m_1, m_2, m_3) \,U_\mathrm{TBM}^T \\
&= \frac{m_1 + m_3}{2} \matrixx{1 & 0 & 0 \\ \cdot & 1 & 0 \\ \cdot & \cdot & 1}  +\frac{m_1 - m_3}{2} \matrixx{1 & 0 & 0 \\ \cdot & 0 & 1 \\ \cdot & \cdot & 0}  + \frac{m_2-m_1}{3}\matrixx{1 & 1 & 1\\ \cdot & 1 & 1 \\ \cdot & \cdot & 1}
\end{split}
\label{eq:TBMmassmatrix}
\end{align}
as long as the masses are different. For degenerate masses we end up with an undefined mixing angle. Since phenomenologically $\dsol \ll \datm$ we take $m_1 = m_2$ as a first approximation:
\begin{align}
 \mathcal{M}_\nu\simeq  \mathcal{M}_\nu^\mathrm{deg} = \matrixx{m_1 & 0 & 0 \\ \cdot & (m_1 + m_3)/2 & (m_1 -m_3)/2 \\ \cdot & \cdot & (m_1 + m_3)/2} .
\end{align}
This matrix can be diagonalized by $\theta_{23}=-\pi/4$, $\theta_{13}=0$ and arbitrary $\theta_{12}$
\begin{align}
  U_{12}^T U_{23}^T \mathcal{M}_\nu^\mathrm{deg}  U_{23} U_{12} = U_{12}^T \,\diag (m_1, m_1, m_3)\, U_{12} = \diag (m_1, m_1, m_3) \,.
\end{align}
Flavor democratic perturbations would then obviously fix $\theta_{12}$ to its TBM value.
Since $U_{12} (\theta_{12})$ is an $SO(2)$ rotation, we find that $\mathcal{M}_\nu^\mathrm{deg}$ is invariant under the $SO(2)$ in flavor space
\begin{align}
 R (\theta_{12}) \equiv U_{23} U_{12} (\theta_{12}) U_{23}^T = \matrixx{c_{12} & s_{12}/\sqrt{2} & s_{12}/\sqrt{2} \\ -s_{12}/\sqrt{2} & \cos^2 (\theta_{12}/2) & - \sin^2 (\theta_{12}/2) \\ -s_{12}/\sqrt{2} & - \sin^2 (\theta_{12}/2) & \cos^2 (\theta_{12}/2)} ,
\end{align}
i.e.~$R^T \mathcal{M}_\nu^\mathrm{deg} R = \mathcal{M}_\nu^\mathrm{deg}$. This can be extended to an $O(2)$, because due to the texture zeroes we also have invariance under the reflection $\diag (-1,1,1)$:
\begin{align}
 \diag (-1,1,1)\, \mathcal{M}_\nu^\mathrm{deg} \, \diag (-1,1,1)  = \mathcal{M}_\nu^\mathrm{deg} \,, && 
U_{23} \,\diag (-1,1,1)\, U_{23}^T = \diag (-1,1,1) \,,
\end{align}
which has the same form in flavor and mass basis. $\mathcal{M}_\nu^\mathrm{deg}$ therefore possesses the $O(2)$ symmetry generated by $R(\theta)$ and $\diag (-1,1,1)$, as well as all their subgroups, which we discussed in the previous section. 

From this discussion it is clear that the chosen $O(2)$ representation $(\nu_e,\,(\nu_\mu+\nu_\tau)/\sqrt{2})^T \sim \vec 2$, $(\nu_\mu-\nu_\tau)/\sqrt{2}\sim \vec 1$, as picked out by the hidden $\Z2$ from Eq.~\eqref{eq:hiddenZ2}, is preferable over other $O(2)$ representations. This is because the doublet representation necessarily results in two degenerate masses, so we should select the smallest $\Delta m_{ij}^2$ for the doublet. Furthermore the subgroup $\mathbb{Z}_2^{\mu\tau}$ fixes $\theta_{13}=0$, so our representation sets all ``small'' neutrino mixing parameters ($\dsol$ and $\theta_{13}$) to zero.\footnote{A connection between the two small parameters is also discussed in Ref.~\cite{Frigerio:2007nn} in a model based on the discrete quaternion group $Q_8$.}

The decomposition of the TBM mass matrix~\eqref{eq:TBMmassmatrix} and the invariance of the different terms under discrete symmetries have been discussed in Refs.~\cite{tripartite} (so-called tripartite model), where the $\Z3$ symmetry~\eqref{eq:Z3} that leads to $\mathcal{M}_\nu^\mathrm{deg}$ was recognized and implemented. The overlying $O(2)$ symmetry has been discussed in Ref.~\cite{scooped}, where a type-II seesaw model with $SO(2)$ symmetry was constructed. Since the symmetry is violated at least by the measured $\dsol\neq 0$, one should however also take perturbations into account to build a viable model. For this reason we devote the next section to a discussion of breaking effects. The model from Sec.~\ref{sec:minimalmodel} will also feature a discussion of the perturbations necessary to accommodate the data from Tab.~\ref{tab:mixingangles}.

%%%%%%%%%%%%%%%%%%%%%%%%%%%%%%%%%%%%%%%%%%%%%%%%%%%%%%%%%%%%%%%%%%%%%%%%%%%%%%%%%%%%%%%%%%%%%%%%%%%%%%%%%%%%%%%%%%%%%%%%

\section{Symmetry Breaking}
\label{sec:symmetrybreaking}
The $O(2)$ symmetry discussed so far can of course not be an exact
symmetry due to the well-measured $\dsol\neq 0$. In this section we
will discuss possible breaking effects, i.e.~perturbations to
$\mathcal{M}_\nu^\mathrm{deg}$. We analyze the most general
perturbations, but first look at the interesting special case of
flavor-democracy as generated for instance by gravity-effects, because they add
exactly a mass term of the form we neglected from
Eq.~\eqref{eq:TBMmassmatrix} to find
$\mathcal{M}_\nu^\mathrm{deg}$. Though there may be other sources
of perturbation that are flavor democratic, we base for definiteness our discussion in
the following subsection on Planck-scale effects. 
We then turn to the general discussion of breaking effects, and how
their effect correlates with the flavor structure of the perturbation
matrices. 
We do not consider contributions to $U_\mathrm{PMNS}$ from the
charged-lepton sector, but these can of course be used to generate
$\theta_{13}$, just like in other TBM models
\cite{HPR}. Renormalization group effects can also lead to sizable~\cite{rgeandtbm}
$\theta_{13}$, but we have nothing new to add to the discussion. We
merely note that in our model it is not possible to generate the right
$\dsol$ and a large $\theta_{13}$ purely through radiative
corrections. As a final remark, in this and the following section
(which deals with an explicit model) we will keep the parameters real,
which simplifies the formulae but apart from that does not lead to reduced physical 
insight in what regards the effect of the perturbations. The main
purpose here is to note the presence of $O(2)$ invariance in the
lepton sector.  

\subsection{Breaking at the Planck Scale}
\label{sec:planckscalebreaking}
The $O(2)$ discussed here is just a global symmetry and will therefore be broken at the Planck scale $M_\mathrm{Pl}$ and maybe even explicitly or spontaneously (which could lead to a dangerous Goldstone boson). Planck scale effects generate the dimension-five Weinberg operator~\cite{Weinberg:1979sa}
\begin{align}
 \mathcal{O}_5 = \frac{C_{\alpha \beta}}{M_\mathrm{Pl} }\,  \left(L_\alpha^T \sigma_2 H \right)\mathcal{C} \left(L_\beta^T \sigma_2 H \right)
= -\frac{C_{\alpha \beta}}{2 M_\mathrm{Pl} }\, \left(L_\alpha^T \mathcal{C}\, \sigma_2 \sigma_j L_\beta \right) \left( H^T \sigma_2 \sigma_j H \right) \,,
\label{eq:planckoperator}
\end{align}
where $\sigma_j$ acts on the $SU(2)_L$ indices of the doublets $H$ and $L_\alpha$, the charge-conjugation matrix $\mathcal{C}$ on spinors and $C_{\alpha\beta}$ describe the flavor-dependent coupling constants. The second equality in Eq.~\eqref{eq:planckoperator} follows from a Fierz identity and can be viewed as a different isospin coupling of the doublets. Gravity is expected to ignore the flavor structure, so one usually assumes $C_{\alpha \beta} = C $, which results in the flavor democratic\footnote{In Ref.~\cite{Berezinsky:2004zb} it is argued that there could be deviations from flavor democracy due to radiative corrections and topological fluctuations (wormhole effects). We will ignore this.} contribution~\cite{Berezinsky:2004zb,planckscale}
\begin{align}
 \delta \mathcal{M}_\nu^\mathrm{Pl} = \mu C \matrixx{ 1 & 1 & 1\\ 1 & 1 & 1\\ 1 & 1 & 1} ,
\label{eq:planck}
\end{align}
with $\mu \simeq v_\mathrm{EW}^2/ M_\mathrm{Pl} \simeq \unit[5 \times 10^{-6}]{eV}$. The above flavor structure of course assumes a diagonal mass matrix of the charged leptons. Since $\mathcal{O}_5$ can only be properly calculated in a theory of quantum gravity, we have no knowledge of the additional coupling constant $C$. It is usually assumed to be $\mathcal{O}(1)$ but we stress that it might differ significantly. For example, using the reduced Planck mass instead of $M_\mathrm{Pl}$ gives $\mu \ra \sqrt{8\pi} \mu$, while from Eq.~\eqref{eq:planckoperator} it is clear that an additional factor of two originates from the $SU(2)_L$ structure of $\mathcal{O}_5$ (as pointed out in Ref.~\cite{Vissani:2003aj}), so the scale of $\delta \mathcal{M}_\nu^\mathrm{Pl}$ is by no means known.
We diagonalize $\mathcal{M}_\nu^\mathrm{deg} + \delta \mathcal{M}_\nu^\mathrm{Pl} $ with the angles $\theta_{23}=-\pi/4$, $\theta_{13}=0$ and $\tan 2 \theta_{12} = \sqrt{8}$, which picks out the TBM solution for the solar mixing angle (just compare $\mathcal{M}_\nu^\mathrm{deg} + \delta \mathcal{M}_\nu^\mathrm{Pl}$ with Eq.~\eqref{eq:TBMmassmatrix}). We stress that TBM can be obtained from the simplest continuous group $SO(2)$ (or $O(2)$) as it is automatically broken to TBM by the flavor-democracy of gravitational effects. Concrete realizations of this symmetry in a UV-complete model will of course turn out to be more complicated, much like models that use discrete symmetries to obtain TBM.

On to the masses: 
the eigenvalue $m_2$ is shifted to $m_1 +3 C \mu$, so $m_1$ and $C \mu$ have to have the same sign. The induced solar mass splitting is given by
\begin{align}
 \Delta m_\mathrm{sol}^2 = (m_1+3 C \mu)^2 - m_1^2 \simeq 6 \,C m_1 \mu \simeq \unit[1.5\times 10^{-6}]{eV^2} C\, \frac{m_1}{\unit[0.05]{eV}} \,,
\end{align}
so we would need $C$ of order $\mathcal{O}(100)$ rather than
$\mathcal{O}(1)$. This could happen via breaking slightly below the Planck scale, e.g.~in extra-dimensional
theories where the ``true'' Planck scale $M_*$ is
reduced~\cite{Berezinsky:2004zb},\footnote{Since we only need $M_*
\simeq M_\mathrm{Pl}/100$, the radius of the extra dimension would
only be $R\simeq \unit[10^{-26}]{mm}$ (even smaller for more than one
extra dimension), which is way outside the presently conducted gravity
tests. Since $M_*$ is still large compared to the electroweak
scale---a fact that we actually employ to make $\delta
\mathcal{M}_\nu$ small---this scheme does not solve the hierarchy
problem.} or via larger than $\mathcal{O}(1)$ coupling constants
$C_{\alpha\beta}$ of $\mathcal{O}_5$.  
Assuming we do not radically boost $\mathcal{O}_5$, we need $m_1 $ and
$m_2 = m_1 + 3 C \mu$ to be large, so this scheme prefers inverted
hierarchy or quasi-degenerate neutrinos. 

While Planck-scale effects naturally break the $O(2)$ symmetry down to
TBM---and might even generate the right $\dsol$---this is no mechanism
to generate a nonzero $\theta_{13}$, because $\delta \mathcal{M}_\nu^\mathrm{Pl}$ is $\mu$--$\tau$ symmetric.

%%%%%%%%%%%%%%%%%%%%%%%%%%%%%%%%%%%%%%%%%%%%%%%%%%%%%%%%%%%%%%%%%%%%%%%%%%%%%%%%%%%%%%%%%%%%%%%%%%%%%%%%%%%%%%%%%%%%%%%%

\subsection{Model Independent Perturbations}
\label{sec:model_independent_perturbations}
Without specifying the origin of the $O(2)$ symmetry, we might as well introduce general perturbations
\begin{align}
 \delta \mathcal{M}_\nu = \matrixx{A & B & B\\ \cdot & C & D\\ \cdot &
\cdot & C} + \matrixx{0 & E & -E\\ \cdot & F & 0\\ \cdot & \cdot & -F}
\equiv \delta \mathcal{M}_\nu^\mathrm{s} + \delta
\mathcal{M}_\nu^\mathrm{as} \, ,
\end{align}
which we have separated into a $\mu$--$\tau$ symmetric ($\delta \mathcal{M}_\nu^\mathrm{s}$) and an antisymmetric ($\delta \mathcal{M}_\nu^\mathrm{as}$) contribution, depending on their behavior under $\mathbb{Z}_2^{\mu\tau}$, as generated by $R(\pi)$ from Eq.~\eqref{eq:mutausymmetrygenerator}. Specifically, $R(\pi) \delta \mathcal{M}_\nu^\mathrm{s} R(\pi) = +\delta \mathcal{M}_\nu^\mathrm{s}$ and $R(\pi) \delta \mathcal{M}_\nu^\mathrm{as} R(\pi) = -\delta \mathcal{M}_\nu^\mathrm{as}$. This decomposition is useful because $\theta_{13}\neq 0$ will only be generated by nonzero $\delta \mathcal{M}_\nu^\mathrm{as}$. Defining the diagonalizing matrix as $U_\nu \simeq U_{23} (-\pi/4) U_{23} (\delta_{23}) U_{13}(\delta_{13}) U_{12} (\theta_{12})$ we find the mixing parameters
\begin{align}
 \delta_{23} \simeq \frac{1}{2}-\sin^2 \theta_{23}\simeq \frac{F}{m_1 - m_3} \,, &&
\delta_{13} \simeq \sin \theta_{13}\simeq  \frac{\sqrt{2} E}{m_1 - m_3} \,, &&
\tan^2 2\theta_{12} \simeq 8 \left(\frac{B}{A-C-D} \right)^2\,,
\label{eq:anglespert}
\end{align}
and the mass splitting
\begin{align}
\dsol \simeq 2 m_1 \sqrt{ (A-C-D)^2 + 8 B^2} \,.
 \label{eq:massespert}
\end{align}
We crosscheck that $\theta_{12}$ is undefined if $\delta \mathcal{M}_\nu$ is of the $O(2)$ symmetric form~\eqref{eq:so2invariant_massmatrix} and leads to TBM if all entries in $\delta \mathcal{M}_\nu$ are equal as in $\delta \mathcal{M}_\nu^\mathrm{Pl}$~\eqref{eq:planck}, as it should be.
For TBM we have $\tan^2 2 \theta_{12} = 8$, so the expression in
parenthesis in Eq.~\eqref{eq:anglespert} should be of order one to yield a
viable mixing angle. The solar mixing parameters are generated by
$\delta \mathcal{M}_\nu^\mathrm{s}$, while $\delta_{13}$ and
$\delta_{23}$ stem from $\delta \mathcal{M}_\nu^\mathrm{as}$, as
stated before. Since we assumed all elements of $\delta
\mathcal{M}_\nu$ to be of similar magnitude in the derivation of
Eqs.~\eqref{eq:anglespert}--\eqref{eq:massespert}, we can still make
the qualitative statement $\dsol \sim m_1 (m_1 - m_3) s_{13}$,
i.e.~the solar mass-squared difference is linear in $s_{13}$.

\begin{figure}[t]
	\begin{center}
		\includegraphics[width=0.48\textwidth]{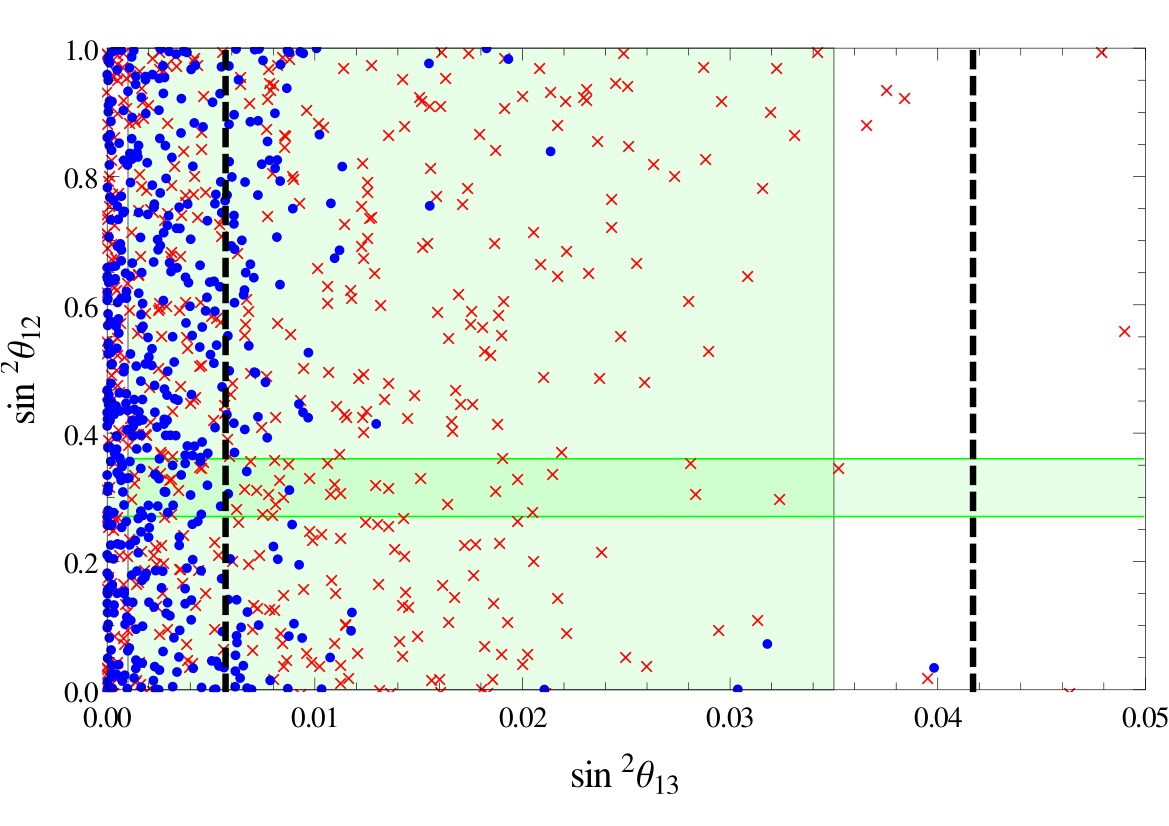}\hspace{3ex}
		\includegraphics[width=0.48\textwidth]{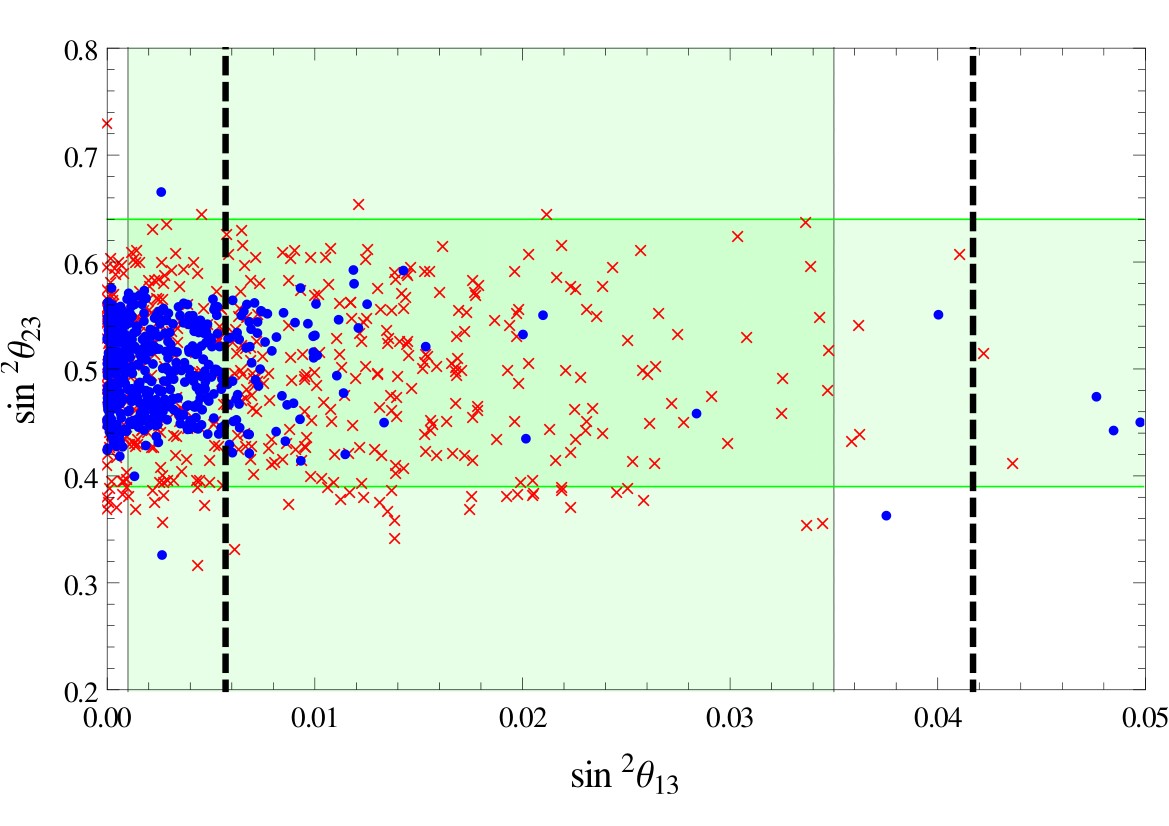}
	\end{center}
		\caption{Scatter plots with arbitrary $O(2)$ breaking perturbations of order $\epsilon = 5\times 10^{-2}$ (blue dots) and $10^{-1}$ (red crosses). The mixing parameters not shown satisfy the $3\sigma$ bounds from Ref.~\cite{theta13}, as do the values inside the green shaded bands. The vertical, dashed black lines give the denote the $95\%$~C.L.~range obtained in Ref.~\cite{Machado} (Eq.~\eqref{eq:t13boundDC}). To not clutter the plot, we only show the NH ranges for $\theta_{13}$, the IH confidence levels vary slightly.}
	\label{fig:scatter}
\end{figure}

To illustrate the points discussed so far, we show some scatter plots
of the mixing angles in Fig.~\ref{fig:scatter}, where we introduced an
arbitrary perturbation matrix to our $O(2)$ invariant mass
matrix~\eqref{eq:so2invariant_massmatrix} with a scale $(\delta
\mathcal{M}_\nu)_{ij} \lesssim \epsilon\, m_\nu$ compared to the
largest neutrino mass $m_\nu$. While $\theta_{23}$ is only slightly
perturbed around its TBM value, $\theta_{12}$ is distributed
approximately uniformly over $[0,\pi/2]$, because for perturbations
larger than those from the Planck scale~\eqref{eq:planck} the value of
$\theta_{12}$ depends crucially on the direction of the perturbation,
i.e.~which entry in $\delta \mathcal{M}_\nu$ is dominant. 

The above formulae~\eqref{eq:anglespert}--\eqref{eq:massespert} suggest $\dsol = 0$ and an undefined $\theta_{12}$ for pure $\mu$--$\tau$ antisymmetric perturbations, which is however only true in linear order. Taking $A=B=C=D=0$ we find the solar mixing parameters for pure $\mu$--$\tau$ antisymmetric perturbations. In terms of deviations from the TBM values $\theta_{23}=-\pi/4$ and $\theta_{13}=0$ we can write
\begin{align}
\dsol\simeq 2 m_1 |m_1 - m_3| \, (\delta_{13}^2 + \delta_{23}^2)  \,, && 
 \sin^2 \theta_{12} \simeq  \begin{cases}
                    \delta_{23}^2/(\delta_{13}^2+\delta_{23}^2) & \mathrm{ NH} \,,\\
                    \delta_{13}^2/(\delta_{13}^2+\delta_{23}^2) & \mathrm{ IH} \,,\\			
                  \end{cases}
\label{eq:theta12fromantisym}
\end{align}
so $\dsol$ is now quadratic in $s_{13}$.
Since we only have two breaking parameters there are relations among the parameters. For NH we can express the deviation from maximal atmospheric mixing via
\begin{align}
 \delta_{23}^2 \simeq \left( \frac{1}{2}- s_{23}^2\right)^2 \simeq \frac{ \Delta m_{21}^2 \, s_{12}^2}{2 m_1 \left(\sqrt{\Delta m_{31}^2 +m_1^2}-m_1 \right)} \,,
\label{eq:smallestneutrinomass}
\end{align}
which can be solved for the smallest neutrino mass $m_1$ as long as $\delta_{23}^2 \Delta m_{31}^2 > s_{12}^2 \Delta m_{21}^2$ is fulfilled, which puts $\theta_{23}$ at the edge of its $2\sigma$ range, making $m_1$ typically large. By varying all parameters in Eq.~\eqref{eq:smallestneutrinomass} over their $3\sigma$ ranges we find a lower bound on $m_1 \gtrsim \unit[0.012]{eV}$. A similar calculation for IH yields the necessary condition  $\delta_{13}^2 \Delta m_{31}^2 < s_{12}^2 \Delta m_{21}^2$ but virtually no lower bound on $m_3$ due to possible finetuned cancellations.

Having discussed the implications of $\mathcal{M}^0:\delta \mathcal{M}^\mathrm{as}:\delta \mathcal{M}^\mathrm{s} = 1:\epsilon:\epsilon$ and $\mathcal{M}^0:\delta \mathcal{M}^\mathrm{as}:\delta \mathcal{M}^\mathrm{s} = 1:\epsilon:0$, we note that the type-I seesaw model of the next section~\ref{sec:minimalmodel} generates the structure $\mathcal{M}^0:\delta \mathcal{M}^\mathrm{as}:\delta \mathcal{M}^\mathrm{s} = 1:\epsilon:\epsilon^2$, which modifies the results from Eq.~\eqref{eq:theta12fromantisym} because both $\delta \mathcal{M}^\mathrm{s}$ and $\delta \mathcal{M}^\mathrm{as}$ contribute to the solar parameters with equal importance, the mixing angle for example can be approximated via
\begin{align}
 \tan^2 2\theta_{12}\simeq   4\, \left(\frac{ \sqrt{2} B + (m_1 - m_3) \delta_{13} \delta_{23}}{A-C-D + (2 m_1 - m_3) (\delta_{13}^2-\delta_{23}^2)}\right)^2 \,.
\label{eq:solarmixinganglefrommodel}
\end{align}
While this softens the behavior discussed below Eq.~\eqref{eq:theta12fromantisym}, we can still find the basic structure of $\delta \mathcal{M}^\mathrm{as}$ in the scatter plots of Fig.~\ref{fig:scatterYukawa}, especially the necessary large deviations from TBM for NH.

%%%%%%%%%%%%%%%%%%%%%%%%%%%%%%%%%%%%%%%%%%%%%%%%%%%%%%%%%%%%%%%%%%%%%%%%%%%%%%%%%%%%%%%%%%%%%%%%%%%%%%%%%%%%%%%%%%%%%%%%

\section{Minimal Model for Type-I Seesaw}
\label{sec:minimalmodel}

To generate three different masses for the charged leptons and not influence the PMNS matrix too much with mixing from the charged leptons, we have to introduce two more Higgs doublets that form an $O(2)$ vector $(\phi_1,\phi_2)\sim \vec{2}$ in addition to the usual Higgs $\phi_3 \sim \vec{1}$~\cite{scooped}.\footnote{Note the different basis for the scalar fields compared to Ref.~\cite{scooped} to emphasize which field combination (here just $\phi_3$) couples to the quarks. Under the reflection $G(0)$: $\phi_1\ra -\phi_1$ and $\phi_2 \ra \phi_2$.} We will first take a look at the neutrino sector.

\subsection{Neutrino Masses}
\label{sec:neutrinomasses}
Putting the right-handed neutrinos into $O(2)$ singlets ($N_R\sim \vec 1 \oplus \vec 1 \oplus \vec 1$), we can write down the invariant terms ($SU(2)_L$ contractions implicit, $\tilde \phi_j \equiv i\sigma_2 \phi_j^*$)
\begin{align}
\begin{split}
 \L \ \supset \ & \frac{1}{\sqrt{2}}(\overline{L}_\mu - \overline{L}_\tau)\, \tilde \phi_3\, (Y_1 N_1 + Y_2 N_2 + Y_3 N_3) \\
&+\left( \overline{L}_e \tilde \phi_1 + \frac{1}{\sqrt{2}} (\overline{L}_\mu + \overline{L}_\tau) \tilde \phi_2\right) (Y_4 N_1 + Y_5 N_2 + Y_6 N_3) \\
&+\left( \overline{L}_e \tilde \phi_2 - \frac{1}{\sqrt{2}} (\overline{L}_\mu + \overline{L}_\tau) \tilde \phi_1\right) (Y_7 N_1 + Y_8 N_2 + Y_9 N_3) +\hc
\end{split}
\label{eq:couplingtosinglets}
\end{align}
In the first line we just coupled the singlets together, while the second uses $\vec{2}\otimes \vec{2}\supset \vec 1$ to build a singlet. The third line uses $\vec{2}\otimes \vec{2}\supset \vec 1'$, so the $Y_7$--$Y_9$ terms are not $O(2)$ invariant. This can be seen by applying the reflection $G(0)$, under which $L_e$ and $\phi_1$ flip signs. We keep these terms for now to see the differences in imposing $O(2)$ and $SO(2)$. Putting everything together, the singlets in Eq.~\eqref{eq:couplingtosinglets} follow from the decomposition $(\vec 2 \oplus \vec 1)^2 \otimes (3\times \vec 1) \supset (6\times \vec 1) \oplus (3\times \vec 1')$,\footnote{This is shorthand for the lengthy decomposition $(\vec 2 \oplus \vec 1) \otimes (\vec 2 \oplus \vec 1) \otimes (\vec 1\oplus \vec 1 \oplus \vec 1) \supset \vec 1\oplus \vec 1\oplus \vec 1\oplus \vec 1\oplus \vec 1\oplus \vec 1\oplus \vec 1'\oplus \vec 1'\oplus \vec 1'$.}
which leads to the Dirac mass matrix
\begin{align}
 m_D = \frac{1}{2}
 \matrixx{\sqrt{2}\, v_1 Y_4 + \sqrt{2}\, v_2 Y_7 & \sqrt{2}\, v_1 Y_5 + \sqrt{2}\, v_2 Y_8& \sqrt{2}\, v_1 Y_6 + \sqrt{2}\, v_2 Y_9\\
  v_2 Y_4-v_1 Y_7 + v_3 Y_1  &  v_2 Y_5 -v_1 Y_8 + v_3 Y_2 & v_2 Y_6-v_1 Y_9 + v_3 Y_3 \\
   v_2 Y_4-v_1 Y_7 - v_3 Y_1  &  v_2 Y_5 -v_1 Y_8 - v_3 Y_2 & v_2 Y_6-v_1 Y_9 - v_3 Y_3 }
\label{eq:diracwithsinglets}
\end{align}
with $\langle \phi_i \rangle = v_i/\sqrt{2}$. Since the $N_i$ are total singlets of the symmetry group, their Majorana mass matrix $\mathcal{M}_R$ is arbitrary and can be taken to be diagonal. Invoking the seesaw mechanism in the symmetry limit $v_{1,2}=0$ we find only one massive neutrino, corresponding to $\nu_\mu - \nu_\tau$. This is obvious because only this $SO(2)$ singlet couples to a right-handed partner and can become massive. While this could be interesting for a normal hierarchy scheme, we would rather build a model that leads to the most general $SO(2)$ symmetric Majorana mass matrix. We will show in the following how this can be accomplished.
We introduce three right-handed neutrinos $N$ in the same $SO(2)$ representation as $L_\ell$. The Majorana mass matrix $\mathcal{M}_R$ is therefore of the form~\eqref{eq:so2invariant_massmatrix}, and since it is in general nonsingular, the inverse $\mathcal{M}_R^{-1}$ is also of the form~\eqref{eq:so2invariant_massmatrix}. The Dirac mass matrix connecting $N$ and $\nu$ stems from the decomposition $(\vec 2 \oplus \vec 1)^3 \supset (4\times\vec 1) \oplus (3\times\vec 1' )$, i.e.~the Lagrangian reads:
\begin{align}
\begin{split}
 \L \ \supset \ & \frac{p_1}{2} (\overline{L}_\mu - \overline{L}_\tau)\, \tilde \phi_3 \,(N_2 - N_3)\\
&+ p_2 \left( \overline{L}_e N_1 + \frac{1}{2} (\overline{L}_\mu + \overline{L}_\tau) (N_2 + N_3) \right) \tilde \phi_3 \\
&+ \frac{p_3}{\sqrt{2}} \left( \overline{L}_e \tilde \phi_1 + \frac{1}{\sqrt{2}} (\overline{L}_\mu + \overline{L}_\tau) \tilde \phi_2\right)  (N_2 - N_3) \\
&+ \frac{ p_4}{\sqrt{2}} (\overline{L}_\mu - \overline{L}_\tau) \left(\tilde \phi_1 N_1 + \tilde \phi_2 \frac{1}{\sqrt{2}} (N_2 + N_3) \right) \\
&+ p_5 \left(  \frac{1}{\sqrt{2}} \overline{L}_e (N_2+N_3) - \frac{1}{\sqrt{2}} (\overline{L}_\mu + \overline{L}_\tau) N_1 \right) \tilde \phi_3 \\
&+ \frac{p_6}{\sqrt{2}} \left( \overline{L}_e \tilde \phi_2 - \frac{1}{\sqrt{2}} (\overline{L}_\mu + \overline{L}_\tau) \tilde \phi_1\right)  (N_2 - N_3) \\
&+ \frac{ p_7}{\sqrt{2}} (\overline{L}_\mu - \overline{L}_\tau) \left(\tilde \phi_2 N_1 - \tilde \phi_1 \frac{1}{\sqrt{2}} (N_2 + N_3)\right) +\hc 
\end{split}
\end{align}
Here $p_5$--$p_7$ denote the $\vec 1'$ terms, which are not $O(2)$ invariant. We once again keep these terms to make the discussion more general. This results in the Dirac matrix
\begin{align}
\begin{split}
 m_D &\equiv m_D^0 (v_3) + \delta m_D (v_{1,2}) =  \frac{v_3}{2\sqrt{2}} 
\matrixx{ 2 p_2 & \sqrt{2} p_5  & \sqrt{2} p_5 \\
-\sqrt{2}  p_5 &  (p_1 + p_2) &  - (p_1 - p_2)  \\
-\sqrt{2}  p_5  &  - (p_1 - p_2)  &  (p_1 + p_2) } \\
&\quad +\frac{1}{2\sqrt{2}} 
\matrixx{ 0 & \sqrt{2} (p_3 v_1 + p_6 v_2 ) & -\sqrt{2} ( p_3 v_1 + p_6 v_2 )\\
\sqrt{2} ( p_4 v_1 + p_7 v_2 ) & -(p_6 + p_7) v_1 + (p_3 + p_4) v_2  & (p_6 -p_7) v_1 - (p_3-p_4) v_2  \\
-\sqrt{2} (p_4 v_1 + p_7 v_2 ) & -(p_6 - p_7) v_1 + (p_3 -p_4) v_2  & (p_6 + p_7) v_1 - (p_3 + p_4) v_2 } .
\end{split}
\label{eq:diracmass}
\end{align}
It is obvious that we obtain an $SO(2)$ symmetric $m_D$~\eqref{eq:invariantmD} for $v_1 = v_2 = 0$, which after seesaw leads to an $SO(2)$ symmetric $\mathcal{M}_\nu$~\eqref{eq:so2invariant_massmatrix} for the light active neutrinos, broken by $\delta m_D (v_{1,2})$.
This generates a nonzero $\dsol$ and deviations from $\theta_{23}=-\pi/4$ and $\theta_{13}=0$, as we will show now. We use the perturbation theory from Sec.~\ref{sec:model_independent_perturbations} for $v_{1,2} \ll v_3$.

We note that $m_D^0 (v_3)$ ($\delta m_D (v_{1,2})$) is symmetric (antisymmetric) under the $\mu$--$\tau$ exchange $\Z2$ $(\nu_\mu,N_2) \leftrightarrow (\nu_\tau,N_3)$, which becomes important when deriving the effects of $\delta m_D$. $\mathcal{M}_R$ and $\mathcal{M}_R^{-1}$ are of course also symmetric under this $\Z2$, due to the $SO(2)$ invariant form.
Using the seesaw formula $m_D \mathcal{M}_R^{-1} m_D^T$ with $m_D = m_D^0 (v_3) + \delta m_D (v_{1,2})$ from Eq.~\eqref{eq:diracmass}, we find in zeroth order of $v_{1,2}$ the general $SO(2)$ invariant Majorana mass matrix~\eqref{eq:so2invariant_massmatrix}. The first order correction $m_D^0 \mathcal{M}_R^{-1} \delta m_D^T +\delta m_D \mathcal{M}_R^{-1}  (m_D^0)^T \propto v_3 v_{1,2}$ is a $\mu$--$\tau$ antisymmetric matrix $\delta \mathcal{M}^\mathrm{as}$ (even--even--odd), which generates $\theta_{13}\neq 0$
\begin{align}
s_{13}^2 \simeq \delta_{13}^2 \simeq \left( \frac{ m_3 (p_2^2 + p_5^2) ( p_3 v_1 + p_6 v_2) + m_1 p_1 ( p_2 p_4 v_1 -p_5 p_7 v_1 + p_4 p_5 v_2 + p_2 p_7 v_2)}{ (m_1 -m_3) p_1 (p_2^2 + p_5^2) v_3}\right)^2 \,,
\label{eq:theta13pert}
\end{align}
and $\delta_{23}^2 = \delta_{13}^2 (v_1 \leftrightarrow -v_2)$.
At order $\delta m_D \mathcal{M}_R^{-1} \delta m_D^T \propto v_{1,2}^2$ we find a $\mu$--$\tau$ symmetric matrix $\delta \mathcal{M}^\mathrm{s}$ (odd--even--odd) that---together with the quadratic contributions from $\delta \mathcal{M}^\mathrm{as}$---gives $\dsol \neq 0$ and fixes the solar mixing angle~\eqref{eq:solarmixinganglefrommodel}. We note that the solar mixing angle takes a simple form if we impose $O(2)$ symmetry ($p_{5,6,7}=0$), depending on the hierarchy and signs of the Yukawa couplings either $\tan \theta_{12}\simeq v_1/v_2$ or $v_2/v_1$.

\begin{figure}[t]
	\begin{center}
		\includegraphics[width=0.48\textwidth]{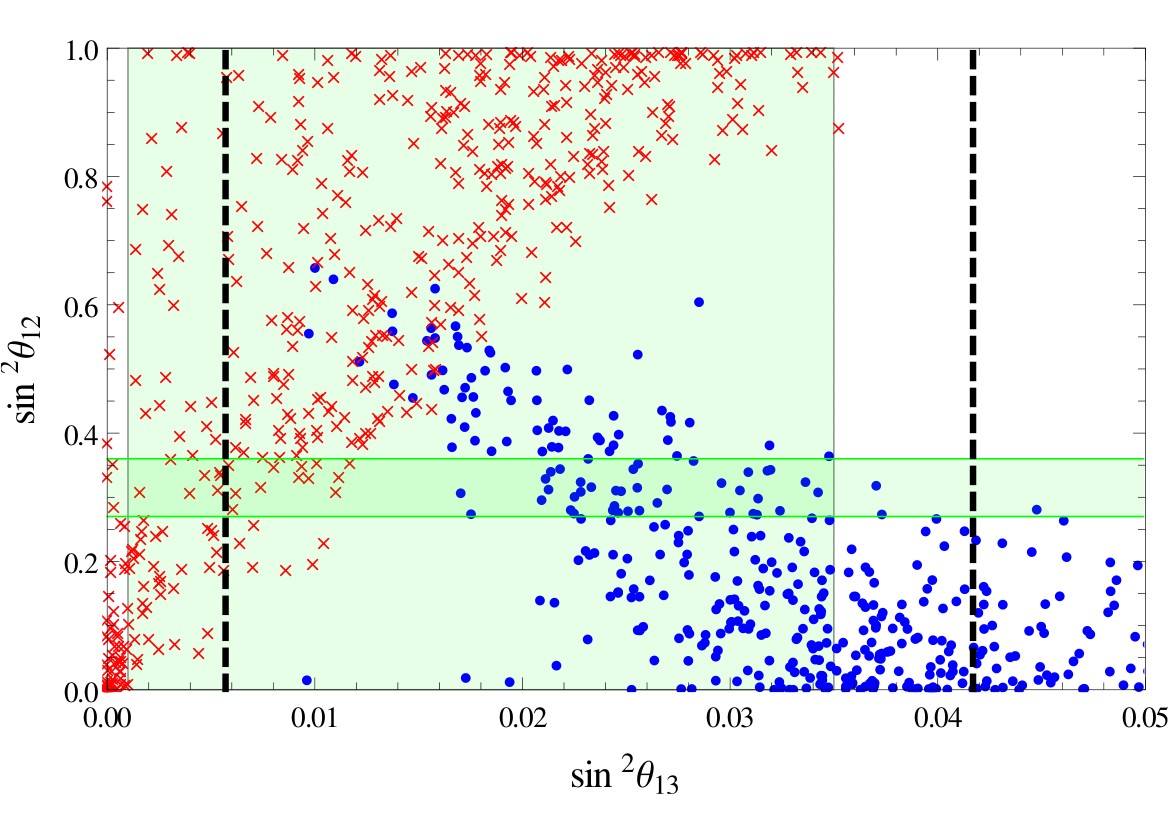}\hspace{3ex}
		\includegraphics[width=0.48\textwidth]{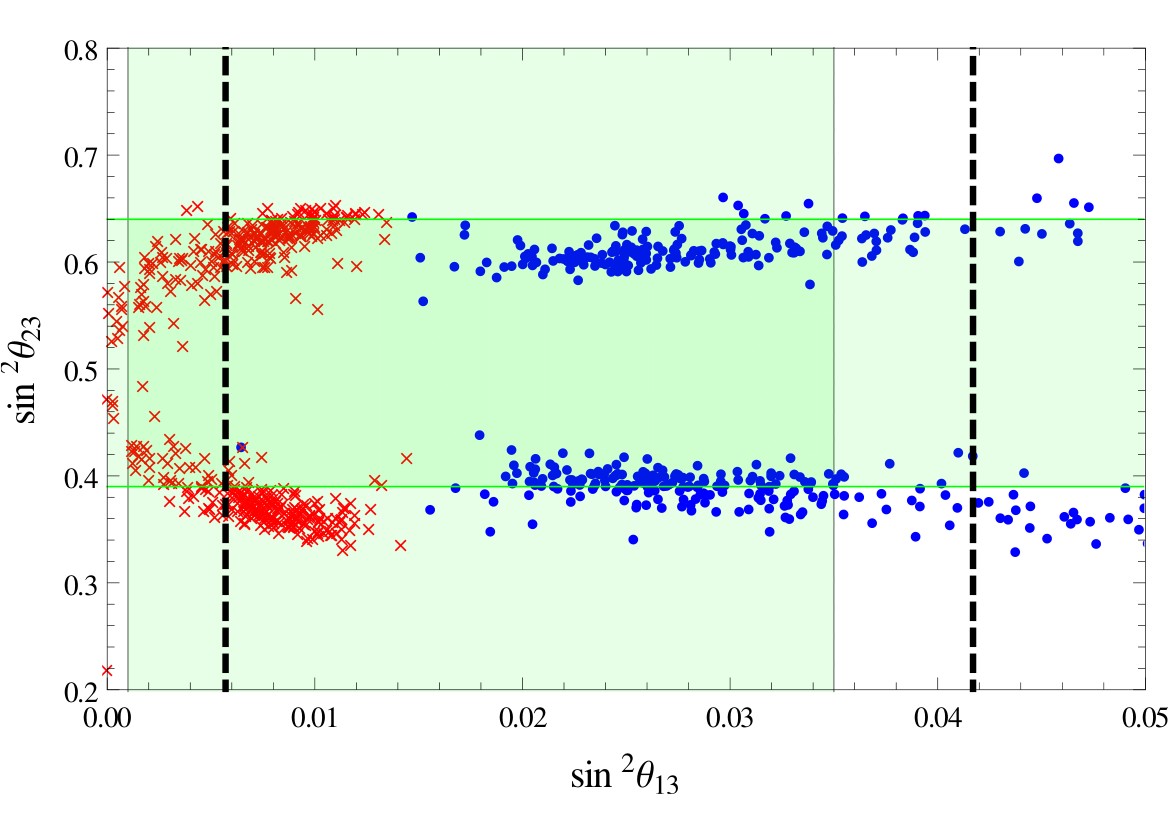}
	\end{center}
		\caption{Scatter plots for the neutrino mixing angles for the type-I seesaw model using perturbations $v_{1,2}/v_3 \sim 10^{-2}$. The blue dots are NH, the red crosses IH solutions. The mixing parameters not shown satisfy the $3\sigma$ bounds from Ref.~\cite{theta13}, as do the values inside the green shaded bands. The vertical, dashed black lines give the denote the $95\%$~C.L.~range obtained in Ref.~\cite{Machado} (Eq.~\eqref{eq:t13boundDC}). To not clutter the plot, we only show the NH ranges for $\theta_{13}$, the IH confidence levels vary slightly.}
	\label{fig:scatterYukawa}
\end{figure}

To illustrate the discussion made so far we show scatter plots for the
mixing angles of $U_\nu$ in Fig.~\ref{fig:scatterYukawa}. Here we
used values $v_{1,2} = \mathcal{O}(1) \,\unit{GeV}$ and real Yukawa
couplings $|p_j| \in  [0.01,3]$, while the largest neutrino mass is
chosen to lie in the range $0.05$--$\unit[1]{eV}$. As already
discussed at the end of
Sec.~\ref{sec:model_independent_perturbations}, the normal hierarchy
solutions demand large deviations from TBM, while inverted hierarchy
is valid even for small $\theta_{13}$.

%%%%%%%%%%%%%%%%%%%%%%%%%%%%%%%%%%%%%%%%%%%%%%%%%%%%%%%%%%%%%%%%%%%%%%%%%%%%%%%%%%%%%%%%%%%%%%%%%%%%%%%%%%%%%%%%%%%%%%%%

\subsection{Charged Leptons}
Taking the right-handed charged leptons in singlets of $O(2)$---$\ell_R \sim \vec 1 \oplus \vec 1 \oplus \vec 1$---results in the mass matrix from Eq.~\eqref{eq:diracwithsinglets}:
\begin{align}
 \mathcal{M}_\ell = \frac{1}{2}
 \matrixx{\sqrt{2}\, v_1 Y_4 + \sqrt{2}\, v_2 Y_7 & \sqrt{2}\, v_1 Y_5 + \sqrt{2}\, v_2 Y_8& \sqrt{2}\, v_1 Y_6 + \sqrt{2}\, v_2 Y_9\\
  v_2 Y_4-v_1 Y_7 + v_3 Y_1  &  v_2 Y_5 -v_1 Y_8 + v_3 Y_2 & v_2 Y_6-v_1 Y_9 + v_3 Y_3 \\
   v_2 Y_4-v_1 Y_7 - v_3 Y_1  &  v_2 Y_5 -v_1 Y_8 - v_3 Y_2 & v_2 Y_6-v_1 Y_9 - v_3 Y_3 } .
\end{align}
Defining the three vectors of Yukawa couplings $\vec{a}\equiv (Y_1, Y_2, Y_3)$, $\vec{b} \equiv (Y_4,Y_5,Y_6)$ and $\vec{c} \equiv (Y_7,Y_8,Y_9)$ (the latter corresponding to the nontrivial $O(2)$ singlets $\vec 1'$), we can write down (assuming real Yukawa couplings and vevs)
\begin{align}
 \mathcal{M}_\ell \mathcal{M}_\ell^\dagger = \matrixx{ \frac{1}{2} (v_1 \vec{b}+ v_2 \vec{c})^2 & \frac{1}{2\sqrt{2}} (v_1 \vec{b} + v_2 \vec{c}) (v_3 \vec{a}+ v_2 \vec{b} - v_1 \vec{c}) & \frac{-1}{2\sqrt{2}} (v_1 \vec{b} + v_2 \vec{c})(v_3 \vec{a} - v_2 \vec{b} + v_1 \vec{c}) \\
\cdot & \frac{1}{4}(v_3 \vec{a} + v_2 \vec{b} - v_1 \vec{c})^2 & \frac{1}{4} \left( (v_2 \vec{b}-v_1 \vec{c})^2 - v_3^2 \vec{a}^2\right) \\
\cdot & \cdot & \frac{1}{4} (v_3 \vec{a} - v_2 \vec{b} + v_1 \vec{c})^2}
\end{align}
or, defining $\vec{A} \equiv  (v_1 \vec{b} + v_2 \vec{c})/\sqrt{2}$, $\vec{B} \equiv ( v_3 \vec{a} + v_2 \vec{b} - v_1 \vec{c})/2$ and $\vec{C} \equiv - (v_3 \vec{a} - v_2 \vec{b} + v_1 \vec{c})/2$:
\begin{align}
  \mathcal{M}_\ell \mathcal{M}_\ell^\dagger = \matrixx{ \vec{A}^2 & \vec{A} \vec{B} & \vec{A} \vec{C}\\ \cdot & \vec{B}^2 &  \vec{B} \vec{C} \\ \cdot & \cdot & \vec{C}^2} .
\end{align}
Since we want to employ the vev hierarchy $v_{1,2} \sim \unit[1]{GeV} \ll v_3 \sim \unit[100]{GeV}$, it is clear that all Yukawa couplings to $\phi_3$, i.e.~$\vec a$, have to be reduced to at least $10^{-2}$ to arrive at the upper lepton mass scale $m_\tau \sim \unit[1]{GeV}$. With the other Yukawa couplings of order one, the vectors $\vec A$, $\vec B$ and $\vec C$ are of similar magnitude, baring cancellations. 
The charged lepton mass matrix is diagonalized using bidiagonalization, i.e.~$U_L^\dagger \mathcal{M}_\ell U_R = \diag(m_e, m_\mu,m_\tau)$, so $U_L$ is the unitary matrix that diagonalizes $ \mathcal{M}_\ell \mathcal{M}_\ell^\dagger$. Since $U_L$ will also contribute to the PMNS matrix $U_\mathrm{PMNS} = U_L^\dagger U_\nu$, we need $U_L\simeq \eins$ since $U_\nu$ obtained in the previous section is already a good approximation to TBM. Consequently, $ \mathcal{M}_\ell \mathcal{M}_\ell^\dagger$ should be almost diagonal, which means the vectors $\vec{A}$, $\vec{B}$ and $\vec{C}$ should be almost orthogonal, with magnitudes roughly $m_e$, $m_\mu$ and $m_\tau$, respectively. Actually, the magnitudes are already sufficient because of the strong hierarchy, so the angles $\alpha_{A B}$, $\alpha_{A C}$ and $\alpha_{B C}$ between the vectors are not that important. Specifically we find roughly
\begin{align}
 U_L \simeq \matrixx{ 1 & c_{A B}\, m_e/m_\mu & c_{A C}\, m_e/m_\tau \\ -c_{A B}\, m_e/m_\mu & 1 & c_{B C}\, m_\mu /m_\tau \\ -c_{A C}\, m_e/m_\tau & -c_{B C}\, m_\mu /m_\tau & 1} ,
\label{eq:ULleptons}
\end{align}
with $c_{I J} \equiv \cos \alpha_{I J}$, which only leads to minor contributions to $U_\mathrm{PMNS}$ even for large angles $\alpha_{I J}$. The dominant effect $s_{23} \ra s_{23} -  c_{23} \, c_{B C} m_\mu/m_\tau$ can be used to soften the strong deviation from maximal mixing predicted by our model for large $\theta_{13}$ (see Fig.~\ref{fig:scatterYukawa}), while $\theta_{13}$ gets shifted to $s_{13}\ra s_{13} - c_{A B} m_e /\sqrt{2} m_\mu$, at most a $10^{-3}$ effect. 

To make $|\vec A|\sim m_e$, we need roughly $|\vec b+\vec c|\sim 10^{-3}$ for $v_1\sim v_2\sim \unit[1]{GeV}$, which is the harshest finetuning in our model. In the limit $v_1 \sim v_2$, $\vec c \sim -\vec b$, the other Yukawa couplings need to satisfy $|v_3 \vec a - 2 v_1 \vec c| \simeq 2 m_\mu$ and $|v_3 \vec a + 2 v_1 \vec c| \simeq 2 m_\tau$, so as expected we need $|\vec a|\sim 10^{-2}$ and $|\vec c|\sim 1$.

As a special case of this model, we can consider the stricter $O(2)$ symmetry instead of $SO(2)$. The neutrino sector barely changes, but we can explicitly determine $\tan \theta_{12}$ as $v_1/v_2$ or $v_2/v_1$, depending on the neutrino hierarchy and other Yukawa couplings. In the charged lepton sector the $O(2)$ symmetry sets $\vec c \equiv 0$, which leads to a massless charged lepton $\sim (v_2,\, -v_1/\sqrt{2},\, -v_1/\sqrt{2})$ and typically large mixing between the two massive ones. The diagonalizing matrix $U_L$ is therefore similar to $U_\nu$ from the neutrino sector, making $U_\mathrm{PMNS} = U_L^\dagger U_\nu $ hard to reconcile with data. 

We can also try different representations for $\ell_R$ for $O(2)$ symmetry. With $\ell_R \sim \vec 1 \oplus \vec 1 \oplus \vec 1'$ or $\ell_R \sim \vec 1 \oplus \vec 1' \oplus \vec 1'$ we find the restriction $\vec a, \vec b \perp \vec c$, which does not lead to a valid lepton sector. The case $\ell_R \sim \vec 1' \oplus \vec 1' \oplus \vec 1'$ only allows for one massive lepton, namely $\sim (v_2,\, -v_1/\sqrt{2},\, -v_1/\sqrt{2})$, which is a bad choice as we can not make this the tauon and therefore have large mixing in $U_L$. Using higher $O(2)$ representations---$\ell_R \sim \vec 2^{(2)} \oplus \vec 1$ or $\ell_R \sim \vec 2^{(2)} \oplus \vec 1'$---is also possible, but typically leads to large mixing in $U_L$ as well.

%%%%%%%%%%%%%%%%%%%%%%%%%%%%%%%%%%%%%%%%%%%%%%%%%%%%%%%%%%%%%%%%%%%%%%%%%%%%%%%%%%%%%%%%%%%%%%%%%%%%%%%%%%%%%%%%%%%%%%%%

\section{Conclusions}
\label{sec:conclusions}

We discussed the unique $O(2)$ symmetry---and its subgroups $SO(2)$,
$\Z{N}$ and $D_N$---connected to the solar mixing angle and $\dsol =
0$, motivated by the phenomenological observation $\dsol \ll
\datm$. Moreover, it generalizes the hidden $\Z2$ symmetry and
leads in addition to $\mu$--$\tau$ symmetry.
The global $O(2)$ is at most an approximate symmetry, as it is
broken at least at the Planck scale. These flavor-democratic
gravitational perturbations pick out the TBM value $\theta_{12} =
\arcsin (1/\sqrt{3})$ and generate $\dsol\neq 0$, however in general
too small unless we modify the Planck scale or coupling a bit. We
stress that the particle-physics Lagrangian can have an $O(2)$
symmetric $\mathcal{M}_\nu$ that is automatically broken to
TBM by flavor-democratic corrections. We constructed a type-I seesaw
model with three Higgs doublets that leads to $SO(2)$ symmetric mass
matrices for the leptons. Breaking the symmetry in the GeV-range can
generate an almost diagonal charged-lepton mass matrix and an
approximately $SO(2)$ symmetric Majorana mass matrix with dominantly
$\mu$--$\tau$ antisymmetric perturbations, so $\theta_{13}$ naturally
receives the largest perturbations. The solar mixing angle
$\theta_{12}$ on the other hand depends on the vevs and several Yukawa
couplings and is almost randomly distributed, so a large $\theta_{12}$
is expected. There are approximate correlations between the mixing
angles that can be tested experimentally.

\begin{acknowledgments}
This work was supported by the ERC under the Starting Grant 
MANITOP and by the DFG in the Transregio 27. J.H.~acknowledges support by the IMPRS-PTFS.
\end{acknowledgments} 

%%%%%%%%%%%%%%%%%%%%%%%%%%%%%%%%%%%%%%%%%%%%%%%%%%%%%%%%%%%%%%%%%%%%%%%%%%%%%%%%%%%%%%%%%%%%%%%%%%%%%%%%%%%%%%%%%%%%%%%%

\appendix

\section{The Group \texorpdfstring{$\boldsymbol{O(2)}$}{O(2)}}
\label{app:o2reps}
In this appendix we briefly discuss the representation theory of the group $O(2)$ of rotations $R(\theta)$ ($\theta \in \mathbb{R}$) and reflections $P$ of the plane (see also Ref.~\cite{Grimus:2008dr}). In the defining two-dimensional representation we find the group action by geometrical considerations\footnote{Another commonly used representation for $P$ is given by $P=\matrixx{0 & 1 \\ 1 & 0}$, which describes a reflection about $(1,1)^T$, compared to $P= \diag (1,-1)$, which is a reflection about the $x_2$ axis.}
\begin{align}
 \2vec{x_1}{x_2}  &\ra R(\theta) \2vec{x_1}{x_2} \equiv \matrixx{\cos \theta & -\sin \theta \\ \sin \theta & \cos \theta} \2vec{x_1}{x_2} , \\
 \2vec{x_1}{x_2} &\ra P \2vec{x_1}{x_2} \equiv \matrixx{1 & 0 \\ 0 & -1} \2vec{x_1}{x_2} \,.
\end{align}
The generators satisfy the group-defining relations
\begin{align} \label{eq:a3}
 R(\theta + 2\pi) = R(\theta)\,, \quad R(\theta_1) R(\theta_2) = R(\theta_1+\theta_2) \,, \quad P^2 = \eins\,, \quad P R(\theta) = R(-\theta) P \,,
\end{align}
where the last relation shows that $O(2)$ is nonabelian. The equivalent definition of 
\begin{align}
 O(2) = \{ M \in \mathbb{R}^{2\times 2} \,|\, M^T M = M M^T = \eins\}
\end{align}
shows that the elements of $O(2)$ have a determinant $\pm 1$. The elements of $O(2)$ with determinant $+1$, i.e.~$R(\theta),\,\theta \in \mathbb{R}$, form the abelian subgroup $SO(2)$. The most general reflection, i.e.~element of $O(2)$ with determinant $-1$, can be written as
\begin{align}
 P (\theta) \equiv R(\theta) P = \matrixx{\cos \theta & \sin \theta \\ \sin \theta & -\cos \theta}
\label{eq:generalreflection}
\end{align}
and we calculate $P (\theta_1) P (\theta_2) = R (\theta_1 - \theta_2)$. Correspondingly, every rotation---and therefore every element of $O(2)$---can be written as a product of reflections (more generally known as the Cartan--Dieudonn\'e theorem).

Besides the trivial representation $\vec 1$, there is another one-dimensional representation $\vec 1'$ generated by $R = 1$ and $P=-1$, so $X \sim \vec 1'$ flips its sign under reflections. There are infinitely many two-dimensional representations $\vec 2^{(n)}$ ($n\in \mathbb{N}$), transforming with multiples of the angle $\theta$, i.e.
\begin{align}
  \2vec{x_1}{x_2}_{\vec 2^{(n)} }  \sim \vec 2^{(n)} &\ra R(n \theta) \2vec{x_1}{x_2}_{\vec 2^{(n)} }  
\end{align}
under rotations. Since we do not make use of different $\vec 2^{(n)}$ in the main part of this paper, we set $\vec 2 \equiv \vec 2^{(1)}$ for convenience.
The tensor product of two two-dimensional representations can be decomposed as follows ($m>n$):
\begin{align}
 \2vec{x_1}{x_2}_{\vec 2^{(n)} } \otimes \2vec{y_1}{y_2}_{\vec 2^{(n)} } &= \left( x_1 y_1 + x_2 y_2\right)_{\vec 1}  \oplus \left( x_1 y_2 - x_2 y_1\right)_{\vec 1'}   \oplus \2vec{ x_1 y_1 - x_2 y_2 }{ x_1 y_2 + x_2 y_1}_{\vec 2^{(2 n)} } ,\\
 \2vec{x_1 }{ x_2}_{\vec 2^{(n)} } \otimes \2vec{y_1  }{ y_2}_{\vec 2^{(m)} } &= \2vec{ x_1 y_1  - x_2 y_2 }{ x_1 y_2 + x_2 y_1 }_{\vec 2^{(m+n)} } \oplus \2vec{ x_1 y_1  + x_2 y_2 }{ x_1 y_2 - x_2 y_1 }_{\vec 2^{(m-n)} } .
\end{align}
Nontrivial tensor products with singlets are given by $\vec 1' \otimes \vec 1' = \vec 1$ and
\begin{align}
 \2vec{ x_1 }{ x_2}_{\vec 2^{(n)} } \otimes (z)_{\vec 1'} = \2vec{- x_2 z }{ x_1 z}_{\vec 2^{(n)} } .
\end{align}
The representation theory for the subgroup $SO(2)$ follows from the above discussion with the remark that $\vec 1 = \vec 1'$, i.e.~there are two singlets.
The finite $O(2)$ subgroups $\Z{N}$ (with elements $[R(2\pi/N)]^m$, $m=0,\dots,N-1$) and $D_N = \Delta (2 N)$ (with elements $[R(2\pi/N)]^m P^n$, $m=0,\dots,N-1$, $n=0,1$) are not of crucial importance for this work, so we omit a detailed discussion. The representation theory for the nonabelian $D_{N\geq 3}$ can be found in Ref.~\cite{discretesymmetries} and Ref.~\cite{Blum:2007jz}.

%%%%%%%%%%%%%%%%%%%%%%%%%%%%%%%%%%%%%%%%%%%%%%%%%%%%%%%%%%%%%%%%%%%%%%%%%%%%%%%%%%%%%%%%%%%%%%%%%%%%%%%%%%%%%%%%%%%%%%%%

\section{Hidden \texorpdfstring{$\Z2$}{Z2}}
\label{app:hiddenZ2}
This appendix is devoted to a short discussion of the so-called hidden $\Z2$, named after its unavoidable appearance in any $\mu$--$\tau$ symmetric Majorana mass matrix (for simplicity assumed to be real). Defining the generator of the $\mu$--$\tau$ interchange symmetry in flavor basis
\begin{align}
 R_{\mu\tau} \equiv \matrixx{1 & 0 & 0 \\ 0 & 0 & 1 \\ 0 & 1 & 0}
\end{align}
we find the symmetric matrix satisfying $[\mathcal{M}_\nu , R_{\mu\tau}] = 0$ in the form
\begin{align}
 \mathcal{M}_\nu = \matrixx{a & b & b\\ \cdot & d & e\\ \cdot &\cdot &d} .
\end{align}
This fixes $\theta_{23}=\pi/4$, $\theta_{13}=0$ and $\tan 2\theta_{12} = 2\sqrt{2}\, b/(d+e-a)$, so we can express one of the entries in $\mathcal{M}_\nu$ via $\theta_{12}$. In terms of physical quantities, $\mathcal{M}_\nu$ then takes the form
\begin{align}
 \mathcal{M}_\nu  =\matrixx{m_1 & 0 & 0 \\ \cdot & (m_1 + m_3)/2 & (m_1-m_3)/2 \\ \cdot & \cdot & (m_1+m_3)/2} + \frac{m_2-m_1}{\sqrt{8}} \ \matrixx{ \sqrt{8} \sin^2 \theta_{12} & \sin 2\theta_{12} & \sin 2 \theta_{12} \\ \cdot & \sqrt{2} \cos^2 \theta_{12} & \sqrt{2} \cos^2 \theta_{12} \\ \cdot & \cdot & \sqrt{2} \cos^2 \theta_{12}} ,
\label{eq:Mnumutau}
\end{align}
which reduces to $\mathcal{M}_\nu^\mathrm{TBM}$ from Eq.~\eqref{eq:TBMmassmatrix} for the value $\theta_{12}^\mathrm{TBM}=\arcsin (1/\sqrt{3})$.
To find the second $\Z2$ symmetry, we first determine the most general symmetric matrix $Z$ with $Z^2 = \mathds{1}$, i.e.~the most general $\Z2$ generator. A little algebra leads to
\begin{align}
  Z (x_1, x_2, x_3) =  \mathds{1} -\frac{2}{x_1^2 + x_2^2 + x_3^2} \matrixx{x_1^2 & x_1 x_2 & x_1 x_3 \\\cdot & x_2^2 & x_2 x_3 \\\cdot & \cdot & x_3^2} .
\end{align}
The invariance condition $[\mathcal{M}_\nu , Z (x_1, x_2, x_3)] = 0$ with $\mathcal{M}_\nu$ from Eq.~\eqref{eq:Mnumutau} then fixes the $x_i$ to the generator of the hidden $\Z2$ in flavor space
\begin{align}
 S(\theta_{12}) \equiv \matrixx{ 
\cos 2\theta_{12} & -\sin 2\theta_{12} /\sqrt{2}  & -\sin 2\theta_{12} /\sqrt{2} \\ 
\cdot & \sin^2 \theta_{12} &  -\cos^2 \theta_{12} \\ 
\cdot & \cdot & \sin^2 \theta_{12}} ,
\end{align}
the other solutions being $Z = R_{\mu\tau}$, $Z = R_{\mu\tau} S(\theta_{12}) = S(\theta_{12}) R_{\mu\tau}$ and of course their negatives $Z\ra -Z$.
We recognize our $G (x)$ from Eq.~\eqref{eq:hiddenZ2} as $G (x) = -S (x) R_{\mu\tau}$.
As special cases we show $S (\theta_{12})$ for the values $\pi/4$ (bimaximal), $\pi/6$ (hexagonal), $\arctan (1/\varphi)$ (golden ratio) and $\arcsin (1/\sqrt{3})$ (tri-bimaximal):
\begin{align}
&\frac{1}{2} \matrixx{ 0 & -\sqrt{2} & -\sqrt{2} \\ \cdot & 1 & -1 \\ \cdot & \cdot & 1} , &
&\frac{1}{4} \matrixx{ 2 & -\sqrt{6} & -\sqrt{6} \\ \cdot & 1 & -3 \\ \cdot & \cdot & 1} , &
&\frac{1}{\sqrt{5}} \matrixx{ 1 &  -\sqrt{2} &  -\sqrt{2} \\ \cdot & 1/\varphi & -\varphi \\ \cdot & \cdot & 1/\varphi} , &
&\frac{1}{3} \matrixx{ 1 & -2 & -2 \\ \cdot & 1 & -2 \\ \cdot & \cdot & 1} , 
\end{align}
with the golden ratio $\varphi = (1+\sqrt{5})/2$. See Refs.~\cite{Rodejohann:2011uz} for a collection of references concerning these symmetries and realizations of these solar mixing angles via discrete symmetries.

Note that the invariance of the Majorana mass matrix under $\Z2$ symmetries is by no means special to the $\mu$--$\tau$ symmetric form. The invariance of any symmetric $3\times 3$ matrix under $\Z2 \times \Z2 \times \Z2$ follows simply from its diagonalizability, as shown in Ref.~\cite{grimus}. For the $\mu$--$\tau$ symmetric case we identified the three $\Z2$ symmetries as generated by $R_{\mu\tau}$, $S(\theta_{12})$ and $S(\theta_{12}) R_{\mu\tau}$.

%%%%%%%%%%%%%%%%%%%%%%%%%%%%%%%%%%%%%%%%%%%%%%%%%%%%%%%%%%%%%%%%%%%%%%%%%%%%%%%%%%%%%%%%%%%%%%%%%%%%%%%%%%%%%%%%%%%%%%%%

\end{document}